\documentstyle[amssymb,preprint,aps]{revtex}

\begin{document}
\author{Rogerio de Sousa$\thanks{%
Present address: Department of Physics, University of Maryland at College
Park, MD 20742-4111. Email: rsousa@physics.umd.edu}$ and G.G. Cabrera}
\address{Instituto de F\'{\i }sica `Gleb Wataghin ', UNICAMP, C.P. 6165,\\
Campinas 13083-970 SP, Brazil}
\title{Precursory Metal-Insulator transition in a small cluster \\
of the `{\em t-J} ' model: Exact analytic results}
\date{8 March 2000}
\maketitle

\begin{abstract}
We study the effect of hole hopping in a doped antiferromagnet described by
the `{\em t-J} ' model, using exact analytic solutions for small clusters.
In spite of the small size, they reveal interesting details about the
magnetic order, which are not apparent in Mean Field treatments or in
numerical calculations. The 4-site cluster with one hole yields the most
interesting physics, displaying different behaviors for the ground state: 
{\em i) }an antiferromagnetic phase for $t\ll J$, where the hole seems to be
localized, not affecting the order of the Heisenberg spins; {\em ii) }%
another regime for $t\sim J$ that presents mixed ferro and antiferro
correlations and coexistence of metallic and insulating behaviors, with the
presence of charge and spin density waves, in what may be the analog of the
spiral phase obtained in Mean Field solutions; and finally {\em iii) }for $%
t\gg J$ , we obtain strong ferromagnetic correlations (maximum spin) and no
density waves, with quantum fluctuations precluding the saturation of the
magnetic moment. This behavior shows traces of a metal-insulator transition
as the hole kinetic energy competes with the antiferromagnetic interactions.
\end{abstract}

The physics of the normal state of High-$T_{c}$ compounds is yet to be
understood after nearly 12 years of great experimental and theoretical
developments. The antiferromagnetic long range order at $T=0$ is rapidly
destroyed when the CuO planes are slightly doped with holes ($x\sim 0.02$
for $La_{2-x}Sr_{x}CuO_{4}$)\cite{keimer}. These holes form $O^{-}$ ions
which coupled to the $Cu^{2+}$ spin 1/2 local moment can be regarded as
singlets centered at the coppers. Looking at this singlet as a hole on a $%
Cu^{2+}$ square lattice, Zhang and Rice \cite{zhang} derived an effective
Hamiltonian, the so called `$t-J$ ' model, which describes the hopping of
holes on a spin lattice, with the spins coupled via an antiferromagnetic
Heisenberg interaction. Mean field theory (MFT) was extensively used by
numerous authors to treat the hole motion on the interacting spin background%
\cite{yoshioda,jayaprakash,kane}. They found an evolution from an antiferro
to a ferromagnetic phase, when the parameter $t/J$ is increased at low or
intermediate hole concentrations. The most interesting feature of those
calculations is the spiral phase that they predict for $t\sim J$, a trait
that seems to have been observed in $La_{2-x}Sr_{x}CuO_{4}$ by neutron
scattering experiments \cite{cheong}. The question then arises on what would
be the effect of quantum fluctuations (which are neglected in MFT) on this
magnetic order, and what is the nature of this spiral state. Since this
phase is induced by hole mobility, it should be accompanied by an
insulator-to-metal transition.

In this paper we calculate exact eigenfunctions for the `$t-J$ ' model
Hamiltonian 
\begin{equation}
H_{t-J}=-t\sum_{{\sigma }{<i,j>}}P_{0}\left( c_{i\sigma }^{{\bf \dagger }%
}c_{j\sigma }+h.c.\right) P_{0}+J\sum_{<i,j>}P_{0}\left( \overrightarrow{S}%
_{i}\cdot \overrightarrow{S}_{j}-\frac{n_{i}n_{j}}{4}\right) P_{0},
\label{htJ}
\end{equation}
where $c_{i\sigma }^{{\bf \dagger }}$ is the fermionic operator that creates
an electron (hole) with spin $\sigma \ $on\ site$\ i$, $n_{i}=\sum_{\sigma
}c_{i\sigma }^{{\bf \dagger }}c_{i\sigma }$ is the number operator, $P_{0}$
projects out double occupied states in the lattice, and the brackets $%
\left\langle ...\right\rangle $ under the summation signs mean nearest
neighbors. The parameter $t$ measures the mobility of carriers which hop
from site to site in an antiferromagnetic Heisenberg background, with
exchange $J>0$. We were able to obtain exact analytic solutions for the
eigenfunctions of \ (\ref{htJ}) in a 4-site cluster, which despite its small
size, shows traces of a metal-insulator transition for the ground state ($%
T=0 $), when the parameter $\lambda =t/J$ is varied. We also propose a way
to analyze the subjacent magnetic order for $\lambda \sim 1$, searching for
a precursory spiral phase in this small cluster. We solve the problems with
one and two holes on a square cluster. The two-hole case is simpler, since
the remaining spins accommodate to form a Resonating-Valence-Bonding state
for small $\lambda $, and holes develop Charge Density Waves (CDW) states.
The other phase ($\lambda \gg 1$) is a spin triplet and the two holes are
delocalized. The physics of the one-hole problem is much richer, since there
is a spin that cannot be compensated in spite of the antiferromagnetic
correlations. As a result, we obtained an intermediate phase, with mixed
insulating and metallic behavior, displaying a variety of magnetic textures,
one of which we identify with the spiral phase predicted by MFT. We are
aware that a small cluster cannot display true phase transitions. However,
broken spin and charge symmetry states can be constructed for the ground
level that, at least qualitatively, can be associated with macroscopic
phases.\cite{falicov}

We first consider the case of two holes on a square cluster. This problem
has dimension $6\times 2^{2}=24$. The ground state energy shows a crossover
at $\lambda =\frac{1}{2(\sqrt{2}-1)}\simeq 1.21$. The small $\lambda $ phase
has an antiferromagnetic character, with vanishing total spin and total $%
S_{z}$ component (note that (\ref{htJ}) commutes with $\sum_{i}S_{i}^{z}$
and $\left( \sum_{i}\overrightarrow{S}_{i}\right) ^{2}$ for periodic
boundary conditions). It is double degenerate and can be written as a
dimerization of singlets on the cluster, as shown in Fig. 1, where the
double links stand for spin singlets and the lone circles for holes. This
resembles closely the Resonating-Valence-Bond state (RVB), as prescribed by
Anderson for the Heisenberg model\cite{anderson2}. Broken symmetry states in
the form of Charge Density Waves (CDW) can be obtained by linear
combinations of the above (for example, $\frac{1}{\sqrt{2}}\left(
|a_{1}>+|a_{2}>\right) $ localizes the holes on one of the diagonals). The
above features are characteristic of the insulating state. In contrast, the
ground state for $\lambda \gtrsim 1.21$ shows a qualitatively different
behavior: it is a spin triplet ($S_{z}=1,0,-1$, see the states $\left|
b_{i}\right\rangle $ below), with admixtures from all the 24 kets of our
base, suggesting some type of disorder induced by hole hopping. They are
given by: 
\begin{eqnarray*}
\left| b_{1}\right\rangle  &=&\frac{1}{4}\left( \left| 
\begin{array}{ll}
0 & 0 \\ 
\uparrow  & \downarrow 
\end{array}
\right\rangle +\left| 
\begin{array}{ll}
0 & 0 \\ 
\downarrow  & \uparrow 
\end{array}
\right\rangle +\left| 
\begin{array}{ll}
\uparrow  & \downarrow  \\ 
0 & 0
\end{array}
\right\rangle +\left| 
\begin{array}{ll}
\downarrow  & \uparrow  \\ 
0 & 0
\end{array}
\right\rangle +\left| 
\begin{array}{ll}
\uparrow  & 0 \\ 
\downarrow  & 0
\end{array}
\right\rangle +\left| 
\begin{array}{ll}
\downarrow  & 0 \\ 
\uparrow  & 0
\end{array}
\right\rangle +\left| 
\begin{array}{ll}
0 & \uparrow  \\ 
0 & \downarrow 
\end{array}
\right\rangle \right)  \\
&&+\frac{1}{4}\left| 
\begin{array}{ll}
0 & \downarrow  \\ 
0 & \uparrow 
\end{array}
\right\rangle +\frac{1}{2\sqrt{2}}\left( \left| 
\begin{array}{ll}
0 & \downarrow  \\ 
\uparrow  & 0
\end{array}
\right\rangle +\left| 
\begin{array}{ll}
0 & \uparrow  \\ 
\downarrow  & 0
\end{array}
\right\rangle +\left| 
\begin{array}{ll}
\uparrow  & 0 \\ 
0 & \downarrow 
\end{array}
\right\rangle +\left| 
\begin{array}{ll}
\downarrow  & 0 \\ 
0 & \uparrow 
\end{array}
\right\rangle \right) \quad , \\
\left| b_{2}\right\rangle  &=&\frac{1}{2\sqrt{2}}\left( \left| 
\begin{array}{ll}
0 & 0 \\ 
\uparrow  & \uparrow 
\end{array}
\right\rangle +\left| 
\begin{array}{ll}
\uparrow  & \uparrow  \\ 
0 & 0
\end{array}
\right\rangle +\left| 
\begin{array}{ll}
\uparrow  & 0 \\ 
\uparrow  & 0
\end{array}
\right\rangle +\left| 
\begin{array}{ll}
0 & \uparrow  \\ 
0 & \uparrow 
\end{array}
\right\rangle \right) +\frac{1}{2}\left( \left| 
\begin{array}{ll}
\uparrow  & 0 \\ 
0 & \uparrow 
\end{array}
\right\rangle +\left| 
\begin{array}{ll}
0 & \uparrow  \\ 
\uparrow  & 0
\end{array}
\right\rangle \right) \quad , \\
\left| b_{3}\right\rangle  &=&\frac{1}{2\sqrt{2}}\left( \left| 
\begin{array}{ll}
0 & 0 \\ 
\downarrow  & \downarrow 
\end{array}
\right\rangle +\left| 
\begin{array}{ll}
\downarrow  & \downarrow  \\ 
0 & 0
\end{array}
\right\rangle +\left| 
\begin{array}{ll}
\downarrow  & 0 \\ 
\downarrow  & 0
\end{array}
\right\rangle +\left| 
\begin{array}{ll}
0 & \downarrow  \\ 
0 & \downarrow 
\end{array}
\right\rangle \right) +\frac{1}{2}\left( \left| 
\begin{array}{ll}
\downarrow  & 0 \\ 
0 & \downarrow 
\end{array}
\right\rangle +\left| 
\begin{array}{ll}
0 & \downarrow  \\ 
\downarrow  & 0
\end{array}
\right\rangle \right) \quad ,
\end{eqnarray*}
with an obvious notation for the basis functions that depicts the geometry
of the square, the symbol $0$ standing for holes. Note that basis kets
entering in one of the states $\left| b_{i}\right\rangle $ do not appear in
any other $\left| b_{j}\right\rangle $, meaning that linear combinations of
these degenerate states will not lead to CDW's (the probability of measuring
a hole in one of the sites is the same for all sites). Therefore the $\left|
b_{i}\right\rangle $ states can be seen as precursors of the metallic
regime. It comes as a surprise that we can see traces of a metal-insulator
transition on such a small cluster.

Now, we turn to the problem of four sites with one hole, which is closer to
the half filled band condition and displays more interesting magnetic
properties. Before obtaining the exact ground state, we propose a
variational solution for the case of low hole mobility ($\lambda \approx 0$%
). This way, we expect to shed light on approximations for larger cluster
problems, since analytic solutions can only be obtained in very special
cases. As shown below, it turns out that our variational state corresponds
to the exact solution when $\lambda $ is smaller than a critical value. For
low$\ t$ (or high magnetic coupling $J$), we expect the magnetic order to
resemble that of the Heisenberg model for a 3-site chain with open boundary
conditions. The ground state for this latter problem has energy $-J$ and is 
\begin{equation}
\left| g_{o}\right\rangle =\frac{1}{\sqrt{6}}\left\{ \left| \uparrow
\uparrow \downarrow \right) -2\left| \uparrow \downarrow \uparrow \right)
+\left| \downarrow \uparrow \uparrow \right) \right\} .  \label{g0}
\end{equation}
Next, we define the state $\left| \phi _{j}\right\rangle $ as having a hole
on the $j$-th site with the magnetic order given by (\ref{g0}) for the
remaining sites, and propose the variational wave function 
\[
\left| \Psi \right\rangle =\sum_{j=1}^{4}\frac{1}{2}\ e^{i\theta _{j}}\left|
\phi _{j}\right\rangle \quad , 
\]
suggesting that the hole motion just introduces a phase $\theta _{j}$ as it
resonates through the cluster. We find the mean energy $E=\left\langle \Psi
\right| H_{t-J}\left| \Psi \right\rangle $ to be 
\[
E=-J+\frac{t}{4}\left[ \cos \left( \theta _{1}-\theta _{2}\right) +\cos
\left( \theta _{2}-\theta _{3}\right) +\cos \left( \theta _{3}-\theta
_{4}\right) +\cos \left( \theta _{1}-\theta _{4}\right) \right] \ , 
\]
which is minimal for $\theta _{1}=0$ , $\theta _{2}=\pi $ , $\theta
_{3}=2\pi $ , $\theta _{4}=3\pi $, with the value $-J-t$. For what range of $%
\lambda $ is this variational solution valid? We will show below that for $%
0<\lambda <\frac{1}{2}$, this is indeed the exact ground state for the
cluster. This state is degenerate with the one obtained by reversing all the
spins, in the negative sector of the magnetization.

To get the exact solution, we note that the Hilbert space has dimension $%
4\times 2^{3}=32$ (the hole can occupy each of the 4 sites and each spin can
be up or down), but the $H_{t-J}$ matrix divides up in 2 blocks of $%
S_{z}=\pm \frac{3}{2}\;$(each of dimension $(4\times 4)$) and 2 of $%
S_{z}=\pm \frac{1}{2}\;$(each of dimension $(12\times 12)$). We only
diagonalize the positive magnetization blocks, the other eigenfunctions
being obtained by flipping the spins of the basis (from now on, we always
work on the positive magnetization sector). In addition, the $\left(
12\times 12\right) $ matrix, when written in the total spin basis, breaks up
in 2 blocks of $8\times 8\;($spin$\;1/2)$ and $4\times 4\;($spin$\;3/2)$.
The energy levels can be seen in Fig. 2 and the eigenstates are given in the
Appendix. The first feature we note in Fig.2 is the great number of level
crossings at $J=2t$. This is the Supersymmetric point, where an exact
solution via the Bethe Ansatz has been obtained in one dimension \cite
{sarkar}, with exact separation of the charge and spin degrees of freedom 
\cite{bares}. We see that the ground state energy undergoes two level
crossovers, with the total spin being 1/2, 1/2 and 3/2, as the parameter $%
\lambda $ increases. This cascade effect on the total spin has already been
reported in numerical calculations\cite{freericks,dagotto}. The first ground
state function holds for $0<\lambda <\frac{1}{2}$, has energy $E=-J-t$, and
it is exactly equal to our variational solution shown above. The hole motion
does not affect the magnetic order, just changes the relative phase between
antiferromagnetic components. Although we expected CDW's, this expectation
is not realized, since there is not enough degeneracy in the ground
manifold. The absence of a broken symmetry state might be ascribed to the
small size of the system. In this case, size effects are lifting the
degeneracy with the first excited level, degeneracy that is restored at the
Supersymmetric point.

The second ground state is double degenerate ($\left| c_{1}\right\rangle $
and $\left| c_{2}\right\rangle $ in the Appendix) and holds for $\frac{1}{2}%
<\lambda <\frac{4+\sqrt{13}}{2}\cong 3.8$, with energy $E=-\frac{J}{2}(1+%
\sqrt{1+12\lambda ^{2}})$. What makes those states non trivial is the fact
that they depend on $\lambda $ (see the functions $a$,$b$, $c$, ... in the
Appendix), making them harder to analyze. Both of them obey a curious rule
of sign when the hole hops, the same pattern repeating with a negative sign,
in the same way as in the variational solution. It is easy to see that we
can build linear combinations of $\left| c_{1}\right\rangle $ and $\left|
c_{2}\right\rangle $ (which are linearly independent but not orthogonal),
with real coefficients, that concentrate the hole on the diagonals, i.e.
CDW's can be obtained with those states. Fig 3 shows the probability $p_{i}$
of finding the hole at the $i$-th site as $\lambda $ varies, for the two
kets $\left| c_{1}\right\rangle $ and $\left| c_{2}\right\rangle $. In the
high mobility limit ($\lambda \rightarrow \infty $), all $p_{i}$\thinspace
's go asymptotically to 1/4, as expected.

As noted in the introduction, MFT's yield a spiral magnetic order for $%
\lambda \sim 1$. We would like to test if our solution resembles this spiral
in any way. We note that the states $\left| c_{1}\right\rangle $ and $\left|
c_{2}\right\rangle $ have no uniform spin distribution. Broken spin symmetry
states in the form of {\em Spin Density Waves} (SDW) can be build with the
above states in the form 
\[
\left| SDW\right\rangle =\cos \left( \frac{\theta }{2}\right) \left|
c_{1}+\right\rangle +\exp \left( i\phi \right) \sin \left( \frac{\theta }{2}%
\right) \left| c_{2}-\right\rangle \ , 
\]
where $\pm $ refer to the manifolds $S_{z}=\pm 1/2$, respectively, and $%
(\theta ,\phi )$ are arbitrary parameters which can be related with the
geometry and pitch of the density wave. As noted before, states $\left|
c_{1}\pm \right\rangle $ and $\left| c_{2}\pm \right\rangle $ are degenerate
and linear independent and were chosen for their simple pattern. Calculation
of the site magnetization $\left\langle SDW\left| \overrightarrow{S}%
_{j}\right| SDW\right\rangle $ shows a spiral spin distribution within the
cluster, with a net ferromagnetic component that precesses with $\phi $, and
whose amplitude is modulated as $\lambda $ and $\theta $ vary. The detailed
calculation yields: 
\begin{eqnarray}
&<&\overrightarrow{S}_{1}>=<\overrightarrow{S}_{3}>=[h(\lambda )\cos \theta
-z(\lambda )]\widehat{z}-g(\lambda )\sin \theta \;\widehat{n}_{\bot }
\label{SDW} \\
&<&\overrightarrow{S}_{2}>=<\overrightarrow{S}_{4}>=[j(\lambda )\cos \theta
+z(\lambda )]\widehat{z}+k(\lambda )\sin \theta \;\widehat{n}_{\bot } 
\nonumber
\end{eqnarray}
where $\widehat{n}_{\bot }=\cos (\phi )\widehat{x}+\sin (\phi )\widehat{y}$
\ is a unit vector on the $xy$-plane and ($h,z,g,j,k$) are functions given
in the Appendix. Fig. 4 depicts this distribution for various choices of $%
(\theta ,\phi )$. Some of the configurations shown closely resemble the
double-spiral state found in Ref. \cite{kane} through a mean-field
calculation. Canting of the spins in doped one-dimensional antiferromagnets
has also been reported previously by one of the authors using a variational
approach\cite{hole}. The presence of CDW and SDW in the above states
suggests that they are precursors of the insulating regime. What is a
remarkable peculiarity of the region $\frac{1}{2}<\lambda <\frac{4+\sqrt{13}%
}{2}$, is that we can also construct a ground state with uniform charge and
spin distributions, {\em i.e.} with the same probability $p_{i}$ of finding
the hole in any site and correspondingly, with the same site magnetization.
We call them Uniform Hole Probability (UHP) states. Interestingly enough,
the only two combinations that yield this result are complex and are only
feasible for $\lambda >1/2$ : 
\begin{equation}
|uhp\ \sigma \rangle =\sqrt{1-u^{2}}\ \overline{\left| c_{1}\ \sigma
\right\rangle }\pm i\text{\/\negthinspace \thinspace }u\ \,\overline{|c_{2}\
\sigma \ \rangle }\;,\;{\rm with}\;u(\lambda )=\frac{1}{6\lambda }\sqrt{%
12\lambda ^{2}-1-\sqrt{1+12\lambda ^{2}}}\;,
\end{equation}
where the $\overline{\left| ...\right\rangle }$ symbol stands to indicate
normalized states and $\sigma =\pm $ refers to the manifolds $S_{z}=\pm 1/2$%
, respectively. For these states, we get 
\[
\begin{array}{c}
p_{1}=p_{2}=p_{3}=p_{4}=\frac{1}{4}\ , \\ 
\\ 
\left\langle S_{z}\left( 1\right) \right\rangle =\left\langle S_{z}\left(
2\right) \right\rangle =\left\langle S_{z}\left( 3\right) \right\rangle
=\left\langle S_{z}\left( 4\right) \right\rangle =\sigma \ \frac{1}{8}\ ,
\end{array}
\]
We can generalize the above state to 
\begin{equation}
|uhp\ \widehat{n}\rangle =\cos \left( \frac{\theta }{2}\right) |uhp+\rangle
+\exp \left( i\phi \right) \sin \left( \frac{\theta }{2}\right) |uhp-\rangle
\;,
\end{equation}
as a state with uniform charge distribution and with the uniform
magnetization pointing along the direction $\widehat{n}$ ($\widehat{n}\,$%
being a unitary vector with latitude $\theta $ and azimuthal angle $\phi $).
Clearly, the $\left| uhp\right\rangle $ kets are precursors of the metallic
state. Thus, in the region $\frac{1}{2}<\lambda <\frac{4+\sqrt{13}}{2}$,
beginning at the Supersymmetric point, the system displays a coexistence of
metallic and insulating behaviors. Two main questions remain: {\em i)} why
is the coexistence region so wide; and {\em ii)} how can we extrapolate
these results to the infinite-size limit? The degeneracy of the ground state
can be ascribed to the vicinity of the Supersymmetric point ($J=2t$) and to
the square symmetry (in a sense, it may be a particular feature of the
4-site problem). The large region spanned by this precursory phase is
certainly a size effect, and it will extrapolate to a single point for the
infinite-size limit, or to a line when the doping is introduced as a
variable.

The third ground state, for $\lambda \gtrsim 3.8$, presents the same
metallic character as in the two-hole problem solved at the beginning of
this paper: it presents ferromagnetic spin correlations but holes are
delocalized, all the base kets having the same admixture, thus avoiding the
saturation of the magnetic moment (this is clearly due to quantum
fluctuations, contrary to what is observed in MFT's). This state is double
degenerate and its energy is $\frac{J}{2}-2t$. CDW's cannot be obtained, as
in the two-hole case.

In conclusion, we described traces of a metal insulator transition in the
ground state as we increase $\lambda =\frac{t}{J}$. Although this treatment
can only be done in small clusters, we were able to do a clear analysis of
the wave function, showing properties that are not apparent in a number of
numerical calculations performed for the same problem\cite{dagotto}. We
analyzed the evolution of the magnetic properties by explicitly writing the
wave functions, in contrast to the usual self-consistent calculation of
magnetic averages. Broken charge and spin symmetry states are proposed as
precursors of macroscopic phases. In particular, we found Spin Density Wave
states that can be identified with the spiral phase obtained by Mean-Field
procedures.

{\bf Acknowledgments. }This work was partially supported by Brazilian CAPES (%
{\em Funda\c{c}\~{a}o Coordena\c{c}\~{a}o de Aperfei\c{c}oamento de Pessoal
de N\'{\i}vel Superior}) and CNPq ({\em Conselho Nacional de Desenvolvimento
Cient\'{\i }fico e Tecnol\'{o}gico}). The authors wish to thank the group of
students associated with Prof. Cabrera for enlightening discussions.\newpage

\subsection{Appendix: Eigenfunctions for the 4-site cluster with 1 hole}

The basis for $S_{z}=\frac{1}{2}$ is (note the order): 
\[
\left\{ 
\begin{array}{c}
\left| 
\begin{array}{ll}
\uparrow & \downarrow \\ 
\uparrow & 0
\end{array}
\right\rangle ,\left| 
\begin{array}{ll}
\downarrow & \uparrow \\ 
\uparrow & 0
\end{array}
\right\rangle ,\left| 
\begin{array}{ll}
\uparrow & \uparrow \\ 
\downarrow & 0
\end{array}
\right\rangle ,\left| 
\begin{array}{ll}
\downarrow & 0 \\ 
\uparrow & \uparrow
\end{array}
\right\rangle ,\left| 
\begin{array}{ll}
\uparrow & 0 \\ 
\downarrow & \uparrow
\end{array}
\right\rangle ,\left| 
\begin{array}{ll}
\uparrow & 0 \\ 
\uparrow & \downarrow
\end{array}
\right\rangle , \\ 
\left| 
\begin{array}{ll}
\uparrow & \uparrow \\ 
0 & \downarrow
\end{array}
\right\rangle ,\left| 
\begin{array}{ll}
\uparrow & \downarrow \\ 
0 & \uparrow
\end{array}
\right\rangle ,\left| 
\begin{array}{ll}
\downarrow & \uparrow \\ 
0 & \uparrow
\end{array}
\right\rangle ,\left| 
\begin{array}{ll}
0 & \uparrow \\ 
\downarrow & \uparrow
\end{array}
\right\rangle ,\left| 
\begin{array}{ll}
0 & \uparrow \\ 
\uparrow & \downarrow
\end{array}
\right\rangle ,\left| 
\begin{array}{ll}
0 & \downarrow \\ 
\uparrow & \uparrow
\end{array}
\right\rangle
\end{array}
\right\} 
\]
and the eigenstates can be classified by their total spin $S$ and total
component $S_{z}$: 
\[
\begin{tabular}{|c|c|}
\hline
$S_{z}=\frac{1}{2}$ & $S=\frac{3}{2}$ \\ \hline
Eigenenergies & Eigenfunctions (not normalized) \\ \hline
$\frac{J}{2}$ & $(1,1,1,0,0,0,0,0,0,-1,-1,-1)$ \\ \hline
$\frac{J}{2}$ & $(0,0,0,1,1,1,-1,-1,-1,0,0,0)$ \\ \hline
$\frac{J}{2}-2t$ & $(1,1,1,1,1,1,1,1,1,1,1,1)$ \\ \hline
$\frac{J}{2}+2t$ & $(1,1,1,-1,-1,-1,-1,-1,-1,1,1,1)$ \\ \hline
\end{tabular}
\]
\strut 
\[
\begin{tabular}{|c|c|}
\hline
$S_{z}=\frac{1}{2}$ & $S=\frac{1}{2}$ \\ \hline
Eigenenergies & Eigenfunctions (not normalized) \\ \hline
$-J-t$ & $(-1,2,-1,1,-2,1,1,-2,1,-1,2,-1)$ \\ \hline
$-J+t$ & $(-1,2,-1,-1,2,-1,-1,2,-1,-1,2,-1)$ \\ \hline
$-t$ & $(-1,0,1,1,0,-1,1,0,-1,-1,0,1)$ \\ \hline
$t$ & $(-1,0,1,-1,0,1,-1,0,1,-1,0,1)$ \\ \hline
$-\frac{J}{2}(1+\sqrt{1+12\lambda ^{2}})$ & \multicolumn{1}{|l|}{$\left|
c_{1}\right\rangle =(-1,1,0,a,b,c,-a,-b,-c,1,-1,0)$} \\ \hline
$-\frac{J}{2}(1+\sqrt{1+12\lambda ^{2}})$ & \multicolumn{1}{|l|}{$\left|
c_{2}\right\rangle =(-1,0,1,-b,2b,-b,b,-2b,b,1,0,-1)$} \\ \hline
$\frac{J}{2}(-1+\sqrt{1+12\lambda ^{2}})$ & $(-1,1,0,-d,-e,-f,d,e,f,1,-1,0)$
\\ \hline
$\frac{J}{2}(-1+\sqrt{1+12\lambda ^{2}})$ & $%
(-1,0,1,e,-2e,e,-e,2e,-e,1,0,-1) $ \\ \hline
\end{tabular}
\]
\newpage

\[
a(\lambda )=\frac{-2+\sqrt{1+12\lambda ^{2}}}{6\lambda }\;;\;b(\lambda )=%
\frac{1+\sqrt{1+12\lambda ^{2}}}{6\lambda }\;;\;c(\lambda )=\frac{1-2\sqrt{%
1+12\lambda ^{2}}}{6\lambda } 
\]
\[
d(\lambda )=\frac{2+\sqrt{1+12\lambda ^{2}}}{6\lambda }\;;\;e(\lambda )=%
\frac{-1+\sqrt{1+12\lambda ^{2}}}{6\lambda }\;;\;f(\lambda )=-\frac{1+2\sqrt{%
1+12\lambda ^{2}}}{6\lambda } 
\]

Base $S_{z}=\frac{3}{2}$: (all eigenfunctions with spin $S=\frac{3}{2}$) 
\[
\;\left\{ \left| 
\begin{array}{ll}
0 & \uparrow \\ 
\uparrow & \uparrow
\end{array}
\right\rangle ,\left| 
\begin{array}{ll}
\uparrow & 0 \\ 
\uparrow & \uparrow
\end{array}
\right\rangle ,\left| 
\begin{array}{ll}
\uparrow & \uparrow \\ 
\uparrow & 0
\end{array}
\right\rangle ,\left| 
\begin{array}{ll}
\uparrow & \uparrow \\ 
0 & \uparrow
\end{array}
\right\rangle \right\} 
\]
\[
\begin{tabular}{|c|c|}
\hline
Eigenenergies & Eigenfunctions \\ \hline
$\frac{1}{2}J$ & $(0,-1,0,1)$ \\ \hline
$\frac{1}{2}J$ & $(-1,0,1,0)$ \\ \hline
$\frac{1}{2}J-2t$ & $(1,1,1,1)$ \\ \hline
$\frac{1}{2}J+2t$ & $(-1,1,-1,1)$ \\ \hline
\end{tabular}
\]

Coefficients for the SDW state given in (\ref{SDW}) 
\[
z(\lambda )=\frac{1+32\lambda ^{2}-%
{\displaystyle {1+8\lambda ^{2} \over \sqrt{1+12\lambda ^{2}}}}%
}{16+256\lambda ^{2}}\ ,\qquad h(\lambda )=\frac{11+128\lambda ^{2}+%
{\displaystyle {5+56\lambda ^{2} \over \sqrt{1+12\lambda ^{2}}}}%
}{48(1+16\lambda ^{2})}\ ,
\]
\[
j(\lambda )=\frac{1+64\lambda ^{2}-%
{\displaystyle {5+56\lambda ^{2} \over \sqrt{1+12\lambda ^{2}}}}%
}{48(1+16\lambda ^{2})}\ ,\qquad g(\lambda )=\frac{2+%
{\displaystyle {2+21\lambda ^{2} \over \sqrt{1+12\lambda ^{2}}}}%
}{6\sqrt{2}\sqrt{1+24\lambda ^{2}+\sqrt{1+12\lambda ^{2}}}}
\]
\[
k(\lambda )=\frac{1+%
{\displaystyle {1+6\lambda ^{2} \over \sqrt{1+12\lambda ^{2}}}}%
}{12\sqrt{2}\sqrt{1+24\lambda ^{2}+\sqrt{1+12\lambda ^{2}}}}
\]

\begin{figure}[tbp]
\caption{ Bi-degenerate ground state for $t/J<1.21$. The double links are
singlets which characterize this state as a RVB insulator. }
\end{figure}

\begin{figure}[tbp]
\caption{ Energy levels of the 4-site cluster with 1 hole. As we increase $%
\lambda$ the ground state energy undergoes two level crossings. Note the
amount of degeneracy for $\lambda=0.5$, the Supersymmetric point. }
\end{figure}

\begin{figure}[tbp]
\caption{ Probability of measuring the hole on the i-{\em th} site; solid
line is for $|c_{1}\rangle $ and dotted for $|c_{2}\rangle $. Those states
span the ground manifold when $1/2<\lambda <3.8$. }
\end{figure}

\begin{figure}[tbp]
\caption{ Spiral magnetic order of SDW states for $t\sim J$. In all cases,
the angle $\phi$ gives the orientation on the $xy$-plane: a) $\theta=\pi/2$%
;~~ b) $\theta = \pi$;~~ c) $\theta = 3\pi/2$. The thin ends of the spins
are meant to point into the page. }
\end{figure}

\end{document}